# HIERARCHY OF FUNDAMENTAL INTERACTIONS IN WAVE UNIVERSE

**A.M. Chechelnitsky,** Laboratory of Theoretical Physics,
Joint Institute for Nuclear Research,
141980 Dubna, Moscow Region, Russia
E'mail: ach@thsun1.jinr.ru

## ABSTRACT

Fundamental interactions (FI) represent the core of physical World Picture, compose the basis of observed in Universe phenomena and flowing in it processes. In the frame of Wave Universe Concept (WU Concept) it is pointed to essential features of observed FI set - its hierarchy, isomorphysm, recurrence character. Hierarchy of $\alpha^{(k)}$ fundamental interactions (FI) is shown by the (infinite) Homological series

$$\alpha^{(k)} = \chi^k \alpha^{(0)} \quad k=...-2, -1, 0, 1, 2, ...,$$

where $\chi$ - Fundamental parameter of hierarchy - Chechelnitsky Number $\chi = 3.66(6)$,
$\alpha^{(0)} = \alpha = e^2/\hbar c$ - Fine Structure constant.
Available experimental data sufficiently confirm the analitical representions of the theory (WU Concept).

## IN SEARCH OF WORLD PICTURE.

Earth and Heaven, Universe, phenomena and its constituted objects - that are invariable subjects of observation by man - Homo Sapiens and Homo Instrumentalis - in the course of the whole of history.

Modern science, essentially, is only continuation of this nonexpire, all swallowing tendency. The actual science differs from the past maybe only by possibility to set up of more comlex, often, grand experiments, by possibility to use of more wide data base and array of accumulated knowledge.

Setting up of more new and new experiments, of course, - is not a self - aim. Beyond all of this invariable tendency remains to create the ordered, well-proportioned World Picture (or, as say ancients - Imago Mundi), among them, and picture of early inaccessible to experiments phenomena of subatomic world. That is aim, undoubtly, worthy of man, great and, possible, completely impracticable.

## COMPONENTS OF MODERN PHYSICAL WORLD PICTURE.
## FUNDAMENTAL INTERACTIONS.

Set of observed objects and phenomena of Universe strikes an imaginations. The modern science, will be say, the Standard Model suggests, that beyond all this brightly and infinite phenomenology the resrict set *only four* fundamental interactions (FI) - *strong, weak, electromagnetic and gravitational* (interactions) stands.

### Nondimensional Constants of Interactions.

According to experimental data, characteristic nondimensional constants of $\alpha_i$ fundamental interactions estimate as follow

$$\alpha_s = e_s^2/\hbar c \sim 1/10 = 0.1$$
$$\alpha_w = e_w^2/\hbar c \sim 1/27$$
$$\alpha_{em} = e_{em}^2/\hbar c \sim 1/137 \div 1/128$$
$$\alpha_g = e_g^2/\hbar c \sim 4.6 \cdot 10^{-40}$$

(for the protons interaction [Perkins, 1991, p. 25]),
where $\hbar = h/2\pi$ - Planck const, e - electron charge, c - light speed.

### Fundamental Questions.

In connection with the physical World Picture, which is described to us by the modern science - physics and cosmology, some natural questions of unbiased reseachers may be arised:

\# Why, according to modern physics representation, *only four* (not less, not more) fundamental interactions (FI) exist?
\# Why charges of known FI have *namely these, not another* values?
\# By what its are motivated?
\# If exists any causal (and analytical) connection between FI of different ranges?
\# Why so extremal *difference* between absolute values of electromagnetic and gravitational FI exists?
Questions may be multipliced...

## WAVE UNIVERSE CONCEPT AND FUNDAMENTAL INTERACTIONS.

### Hierarchycal Structure of Universe.

One of most bright, evident phenomenon of real Universe is its *hierarchycal structure*. In Universe obviously *elementar object of matter* (EOM) of different scale - atoms, stars, galaxies, etc. are observing.

### Matter Levels. ″Wave Universe Staircase″ of Matter.

That incontrovertible fact takes inself natural reflection in the *Wave Universe concept* [Checheknitsky, 1980-1997]. Hierarchy of $U^{(k)}$, k=...-2.-1,0,1,2,... matter Levels form the ″Wave Universe Staircase″ of matter. In the analitical, mathematical plane this hierarchy may be represented by *Homological series* (GS) of characteristic parameters EOM of each Level of matter.

### Layers of Matter.

It is evident fact too, that not all Levels of matter equally brightly represent in observations. In that sense most brightly in real Universe its are manifested only some clasters of close situated matter Levels, it will be say, *Layers of matter.* Most characteristic from its are

# *Atomic (subatomic) Layer* of matter. It contain some Levels of matter, connected with populations of atoms, particles, etc.

# *Stellar (Star) Layer* of matter. It contain some matter *Levels*, connected with populations of stars, planets, etc.

# *Galactical Layer* of matter. Objects of matter - different galaxies, etc.

### Fundamental Parameter of Hierarchy.

At 70-th in investigation of wave structure of Solar system [Chechelnitsky, 1980] it have been discovered significant arguments for existance of *Shell structure, hierarchy and similarity - dynamical isomorphysm -* of Solar system Shells.

First of all, that concerned to dynamical isomorphysm of clearly observed $G^{[1]}$ and $G^{[2]}$ Shells, connecting respectively with I (Earth's) and II (Jovian) groups of planets.

It was determined that arrangement of physically distinguished - *elite* (particularly powerful, strong - *dominant*) orbits of Mercury in $G^{[1]}$ (and Jupiter in $G^{[2]}$), Venus in $G^{[1]}$ (and Saturn in $G^{[2]}$) Shells brightly underline the similarity of geometry and dynamics of processes, flowing in these Shells, with accuracy up to the some scale factor.

As the quantitative characteristics of that isomorphism, the recalculation coefficient $\chi$ - *Fundamental parameter of hierarchy (FPH)* - may be used the ratio, for instance, of

# (Keplerian) orbital velocities v

$v_{ME}/v_J = 47.8721 \text{ km} \cdot \text{s}^{-1}/13.0581 \text{ km} \cdot \text{s}^{-1} = 3.66608 \Rightarrow \chi$,

$v_V/v_{SA} = 35.0206 \text{ km} \cdot \text{s}^{-1}/9.6519 \text{ km} \cdot \text{s}^{-1} = 3.62836 \Rightarrow \chi$,

# Sectorial velocities L

$L_J/L_{ME} = 1.01632 \cdot 10^{10} \text{ km}^2 \cdot \text{s}^{-1}/0.27722 \cdot 10^{10} \text{ km}^2 \cdot \text{s}^{-1} = 3.66608 \Rightarrow \chi$,

$L_{SA}/L_V = 1.37498 \cdot 10^{10} \text{ km}^2 \cdot \text{s}^{-1}/0.37895 \text{ km}^2 \cdot \text{s}^{-1} = 3.628357 \Rightarrow \chi$,

# Semi-major axes a

$a_J/a_{ME} = 5.202655 \text{ AU}/0.387097 \text{ AU} = 13.440164 = (3.666082)^2 \Rightarrow \chi^2$,

$a_{SA}/a_V = 9.522688 \text{ AU}/0.723335 \text{ AU} = 13.164975 = (3.628357)^2 \Rightarrow \chi^2$,

# Orbital periods T (d - days)

$T_J/T_{ME} = 4334.47015 \text{ d}/87.96892 \text{ d} = 49.272744 = (3.666082)^3 \Rightarrow \chi^3$,

$T_{SA}/T_V = 10733.41227 \text{ d}/224.70246 \text{ d} = 47.76722 = (3.6283568)^3 \Rightarrow \chi^3$.

In the published at 1980 monograph [Chechelnitsky,1980] (date of manuscript acception - 11 May 1978) this dynamical isomorphysm, similarity of geometry and dynamics of physically distinguished orbits of I (Earth's) and II (Jovian) groups were analized.

According to the content of "Heuristic Analysis" division [Chechelnitsky, 1980, pp.258-263, Fig.17,18] similarity coefficient - recalculation scale coefficient of megaquants

$$D^I = L_{ME}/3 = 0.924 \cdot 10^9 \text{ km}^2 \cdot \text{s}^{-1}$$

$$D^{II} = L_J/3 = 3.388 \cdot 10^9 \text{ km}^2 \cdot \text{s}^{-1}$$

of L - sectorial velocities (actions, circulations) of I and II groups of planets is equal

$$D^{II}/D^I = L_J/L_{ME} = 3.66(6) \Rightarrow \chi.$$

It was not surprise, that transition to another Shells of Solar (planetary) system (to Trans-Pluto and Intra-Mercurian Shells) would be characterized with the same $\chi$ - *Fundamental parameter of hierarchy* (FPH) $\chi = 3.66(6)$.

**Universality of FPH.**

Analysis of (mega) wave structure of physically autonomous satellite systems of Jupiter, Saturn, etc., indicated, that discovered $\chi$ Fundamental parameter of hierarchy (FPH) plays in its the similar essential role, as in the Solar (planetary) system, characterizing the hierarchy, recursion and isomorphysm of Shells.

Thus, it takes shape the essentially *universal* character of (FPH) - its validity for the analysis of (mega) wave structure *of any WDS*.

That corresponds to representations, connected with *co-dimension principle* [Chechelnitsky, 1980, p.245]:

"...fundanental fact is that when we pass on to another WDS, the value of d⁻ [character value of sectorial velocity (action, circulation)] doesn't remain constant, but varies according scales of these systems. This fact is the consequence of *co-dimension principle*..."

**"Magic Number" ("Chechelnitsky Number", FPH)  $\chi = 3.66(6)$.**
**Role and Status of Fundamental Parameter of Hierarchy in Universe.**

Previous after primary publications [Chechelnitsky, 1980-1985] time and new investigations to the full extent *convince* the theory expectations, in particular, connected with the $G^{[s]}$ Shells hierarchy in each of such WDS, with the hierarchy of Levels of matter (and WDS) in Universe, with the exceptional role of the *introduced in the theory* c FPH [Chechelnitsky, (1978) 1980-1986].

The very brief resume of some aspects of these investigations may be formulated in frame of following short suggestion.

**Proposition (Role and Status of c FPH in Universe)** [Chechelnitsky, (1978) 1980-1986]

\# The central parameter, which organizes and orders the dynamical and physycal structure, geometry, hierarchy of Universe

  ∗ "*Wave Universe Staircase*" of matter Levels,

  ∗ *Internal* structure each of real systems - wave dynamic systems (WDS) at *any Levels* of matter, is (manifested oneself) $\chi$ - the *Fundamental Parameter Hierarchy (FPH)* - nondimensional number $\chi = 3.66(6)$.

\# It may be expected, that investigations, can show in the full scale, that $\chi$ - FPH, generally speeking, presents and appears *everywhere* - in any case, - in an extremely wide circle of dynamical relations, which reflect the geometry, dynamical structure, hierarchy of real systems of Universe.

We aren't be able now and at once to appear all well-known to us relations and multiple links, in which one self the [Chechelnitsky] $\chi = 3.66(6)$ *"Magic Number"* manifests.

We hope that all this stands (becomes) possible in due time and with new opening opportunities for the publications and communications.

**HIERARCHY OF MATTER LEVELS AND FUNDAMENTAL INTERACTIONS.**

In the frame of Wave Universe Concept (WU Concept) it may be suggested that hierarchy of $\alpha^{(k)}$ fundamental interactions (FI) corresponds to hierarchy of $U^{(k)}$, k=...-2,-1,0,1,2,... matter Levels. In particular, cluster of neighbouring, fundamental interactions corresponds to Atomic (Subatomic) Layer of matter - its some matter Levels. Among its the well-known in modern physics *strong, weak, electromagnetic interactions* are most brightly manifested.

**Hierarchy of Fundamental Interactions.**
  **Proposition ($\alpha$ Hierarchy - $\alpha$ Homology).**
  Set of observed interactions, phenomena, objects dynamical structures of Universe
  \# Connects with infinite hierarchy $U^{(k)}$, k=...-2.-1,0,1,2,... matter Levels,
  \# Connects and is defined by the infinite hierarchy of Fundamental Interactions (FI)
$$\alpha^{(k)}, \quad k=...-2.-1,0,1,2,...,$$
it will be say, by *$\alpha$ Hierarchy* of FI
  \# $\alpha$ Hierarchy, genarally say, is *infinite*,
  \# This $\alpha$ Hierarchy is represented by Homological series of FI (by the *$\alpha$ Homology*) in form
$$\alpha^{(k)} = \chi^{k}\alpha^{(0)}, \quad k=...-2.-1,0,1,2,...$$

As *prime image (eponim)* it is reason to choose the *Fine Structure Constant* (FSC) - nondimensional constant of electromagnetic interaction
$$\alpha^{(0)} = \alpha_{em} = \alpha = e^2/\hbar c$$
where e - electron charge, $\hbar$ - Planck constant, c - light speed.

\# Well-known from Standard Model - strong, weak, electromagnetic, gravitational FI belong to $\alpha$ Hierarchy and its are represented by $\alpha$ Homology in form

*Strong FI*
$$\alpha^{(2)} = \alpha_s = e_s^2/\hbar c = \chi^2\alpha^{(0)} = \chi^2\alpha = \chi^2 e^2/\hbar c = 0.0981,$$

*Weak FI*
$$\alpha^{(1)} = \alpha_w = e_w^2/\hbar c = \chi\alpha^{(0)} = \chi\alpha = \chi e^2/\hbar c = 0.02675,$$

*Electromagnetic FI*
$$\alpha^{(0)} = \alpha_{em} = \alpha = e^2/\hbar c = 1/137.036 = 0.007297$$

and - for the one of near arrangeing FI in the *Gravitational Layer* of matter (for the electrons interaction) - *Gravitational FI*

$$\alpha^{(-75)} = \alpha_g^{(-75)} = (e_g^{(-75)})^2/\hbar c = \chi^{-75}\alpha = \chi^{-75}e^2/\hbar c = 0.348986 \cdot 10^{-44}$$

## THEORY AND EXPERIMENT.

### Stationary Values of FI Constants.

In the world physical literature variety of experimental estimations of $\alpha^{(k)}$ FI values is circulated. We take (spare) the special attention to those of its, which describe some asimptotical, limitational (apparently, convergent, fixed, stable) its states.

For the definitness it would be later named these experimental estimations - as *stationary* (values).

## STRONG INTERACTION.

### Constant of Strong FI.

The developed now experimental situation most brightly and evidently represents the well-known Figure of [RPP, 1997, Fig.9.2 in Division "Lattice QCD"]. It is not difficult to point, that observed in many experiments asimptotical, limitational, apperently, stationary value of $\alpha_s$ FI constant lie at region

$$\alpha_s \approx 0.098 \div 0.1$$

Let cite some $\alpha_s$ values - result of concrete experiments, - quite corresponding to predictions of theory (WU Concept). References take from RPP

# RPP, p.79: "...The result can be combined to give
$$\alpha_s(M_z) = 0.112 \pm 0.002 \pm 0.04...$$
# RPP, p.80: "...A fit to $\Upsilon$, $\Upsilon'$ and $\Upsilon''$ gives
$$\alpha_s(M_z) = 0.108 \pm 0.001 \text{ (expt.)}$$
# RPP, p. 81: "... the Standard Model is used
$$\alpha_s(M_z) = 0.104$$
# RPP, p. 90: $\quad\quad\quad\quad\alpha_s(M_z) = 0.101 \pm 0.008$
# RPP, p.90: "...Nonsuper symmetric unified theories predict the low value
$$\alpha_s(M_z) = 0.073 \pm 0.001 \pm 0.001$$
# RPP, p. 91: $\quad\quad\quad\quad\alpha_s = 0.101(8)$
# RPP, p.105: $\quad\quad\quad\quad\alpha_s = 0.103 \pm 0.008$

It is possible another, not less effective way to the determination of the $\alpha_s$ FI constant.

### Electron - Positron Annigilation to Adrons and Lepton - Lepton Decay.
### Dynamical Isomorphysm.

Examination of process of $e^+e^-$ annigilation in hadrons

$$e^+e^- \to \text{hadrons}$$

is possible with idea of it "close analogy" [see, for instance, Perkins, 1991, p.271] with process of lepton-lepton decay

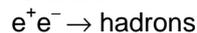
$$e^+e^- \to \mu^+\mu^-$$

The characteristical, definitable parameter - relation of total cross sections

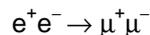
$$R = \sigma(e^+e^- \to \text{hadrons})/\sigma(e^+e^- \to \mu^+\mu^-),$$

as result of many experiments on high energy $e^+e^-$ collaiders, is practically stationary at E>10 Gev. This "...confirms the point character of $e^+e^-$ adrons process, happened analogous to $e^+e^- \to \mu^+\mu^-$ process..." [Perkins, 1991, p.255].

The experimental value of R is equal

$$R = \sigma(e^+e^- \to \text{hadrons})/\sigma(e^+e^- \to \mu^+\mu^-) = 11/3.$$

Therefore, it may be assumed the some dynamical isomorphism (similarity, "close analogy") of these two processes with accuracy to similarity parameter (recalculation parameter) R=11/3.

### Strong Interaction and (FPH) Constant

It is not difficult to comprehend, that experimentally obtaining value of R=11/3=3.66(6) similarity parameter, in reality, completly coincides with value of $\chi=3.66(6)$ - Fundamental parameter of hierarchy

$$R = 11/3 = 3/66(6) \Rightarrow \chi = 3.66(6).$$

Therefore, it may be justified the relation

$$R = \sigma(e^+e^- \to \text{adrons})/\sigma(e^+e^- \to \mu^+\mu^-) = \chi = \alpha_s/\alpha_w = e_s^2/e_w^2.$$

Experiment spontaneous, directly and confidently fixed and confirm following from the theory (WU Concept) value of $\alpha_s$ strong and $\alpha_w$ weak FI relation.

**WEAK INTERACTION.**

**Constant of Weak FI.**

It is observed comparatively wide field of $\alpha_w$ weak FI estimations. Apparently, this connect with situation when point some "intermediate", not asimptotical evaluations, in more degree depending from dynamical circumstan cies of experiment (transfer momentum, etc).

**Influence of Model Representations.**

It may be suppose that incertainty in $\alpha_w$ estimations is connected also with nonjustified propositions, limitations and links of developed theoretical models of weak FI (for instance, Weinberg-Salam model), in frame of which the experimental data are calculated and comprehended. This interesting aspect deserves the special deep discussion.

**Weak Angle of Mixing. Comparision with Experiment.**

Let take as definition the following representation for the $\theta_w$ - *Weak angle of mixing* (Weinberg angle)

$$e/e_w = \sin\theta_w$$

It is wide used in Standard Model (Weinberg - Salam model), and $\theta_w$ - for the comparison of theory and experimental data. In frame of discussed WU Concept representations it is not difficult to receive numerical, theoretical representation for this angle. From the comparision

$$e_w^2/e^2 = \sin^{-2}\theta_w \Rightarrow \chi = 3.66(6)$$

it is followed

$$\sin^2\theta_w = 1/\chi = 0.272(272),$$
$$\sin\theta_w = 0.522(232), \theta_w = 31°.48.$$

As the case of preliminary reason for reflection let take only one reference [Perkins, 1991]. In division "Assimetries of polarized electrons decay on deuterons", it is constated [p.320]: "...The final result has the form

$$\sin^2\theta_w = 0.22 \pm 0.02$$

and is coordinated with previous estimations. However, if save the $\rho$ relation of neutral and charged currents as a free parameter, then result will be another:

$$\rho = 1.74 \pm 0.033,$$
$$\sin^2\theta_w = 0.293 \pm 0.033 \pm 0.100,$$

because $\rho$ and $\sin^2\theta_w$ values are strong correlated...".

From one side, it is not difficult to point the comporativeness of experimental data

$$\sin^2\theta_w = 0.22 \div 0.293$$

and WU Concept result

$$\sin^2\theta_w = 0.272.$$

From another side, these data of experiments reflect still surviving inconsistency, sqeezed in the interpetation of experimental data. It seems, that results of experiments highly nonwillingly invide in imperatives of Standard Models - in restricting relations of developed theory. Problem is so interesting and entertaining that we intend to return to it with detailed critical analysis of developed sutiation.

**Weak Interaction and $\chi$ FPH.**

The Fundamental parameter hierarchy ($\chi$ FPH) plays decisive role in hierarchy of weak and electromagnetic FI. To be convinced in it, let cite data of experiment, which point another way for experimental definition of $\alpha_w$ constant of weak FI.

**Dynamical Isomorphysm of Electron and Neutrino Decay (at Nucleons).**

Experiments with leptons (electrons and neutrino) decay on nucleons are characterized by standard functions of nucleons

$F_1^{eN}, F_2^{eN}$ - for electromagnetic,
$F_1^{\nu N}, F_2^{\nu N}, F_3^{\nu N}$ - for weak interactions.

Accumulated information leads, in particular, to relation

$$F_2^{\nu N} \leq (18/5) F_2^{eN}$$

Commentary [Perkins,1991, Fig.8.11] sounds as follow: "...This is the first comparison of $F_2^{\nu N}$ function, mesured by *neutrino* - nucleon decay in CERN neutrino bearn..., with the SLAC data of $F_2^{eN}$ function in *electron* - nucleon decay with the same $q^2$... Both data sets coincide each to other, if points of *electron* decay are multipliced by 18/5..."

Thus, it is observed the dynamical isomorphysm (similarity) of weak and electromagnetic decay processes with accuracy to similarity parameter

$$F_2^{\nu N}/F_2^{eN} = r \cong 18/5$$

It is not difficult to comprehand, that these orienting data of experiment, in reality, correspond to the $\chi$ FPH

$$r \cong 18/5 = 3.6 \Rightarrow 3.66(6).$$

Another words, experimental data directly and confidently fixed the weak and electromagnetic constant FI relation

$$F_2^{\nu N}/F_2^{eN} \cong \chi = 3.66(6) = \alpha_w/\alpha_{em} = e_w^2/e^2,$$

characterized by FPH $\chi=3.66(6)$.

## ELECTROMAGNETIC INTERACTION.

### Constant of Electromagnetic FI.

Experimental value of $\alpha_{em}$ FI in the HEP (high energy physis) is estimated as lieing in region

$$\alpha_{em} \sim 1/137 \div 1/128$$

in dependence of transfering momentum (with growth of transfering momentum - the $\alpha_{em}$ grows - in difference of $\alpha_{em}$ and $\alpha_w$).

It is natural to consider that the well-known from QED and macroworld value of electromagnetic FI constant is fixed, stationary and is equal to

$$\alpha_{em} = \alpha = e^2/\hbar c = 1/137.036,$$

where $\alpha = e^2/\hbar c$ - Fine Structure constant,
$\hbar = h/2\pi$ - Planck const, e - electron charge, c - light speed.

## GRAVITATIONAL INTERACTION.

### Constant of Gravitational FI.

Existing estimations of $\alpha_g$ gravitational FI constant scarcely may be considered as based on *specially* created experiments. So, accepted now (indirect) experimental estimation point the value (for the *protons* interaction - Perkins, 1991, p.25)

$$\alpha_g \sim 4.6 \cdot 10^{-40}.$$

In frame of WU Concept we point (more definitely) the following value of $\alpha_g$ gravitational FI constant (for the *electrons* interaction)

$$\alpha^{(-75)} = \alpha_g^{(-75)} = (e_g^{(-75)})^2/\hbar c = \chi^{-75}\alpha = \chi^{-75}e^2/\hbar c = 3.4898 \cdot 10^{-45}.$$

It corresponds to one of matter Levels, it will be say, of *Gravitational Layer* of matter.

This is claster of near lieing Levels of matter. Of course, the "force" of gravitational FI, corresponding to one of matter Levels of Gravitational Layer of matter, is very small in comparision to "forces" of strong, weak and electromagnetic FI.

### Gravitational FI - Some Additional Aspects.

The base for that representation for $\alpha_g$ gravitational FI constant is the following assumption of WU Concept

**Proposition (Gravitation and Electromagnetism).**

\# Gravitation and Electromagnetism (as and another FI) in Wave Universe passess by fundamental *stable wave link*, characterized, in particular, by properties of *commensurability, stable resonance*.

\# Fundamental constants of gravitation and electromagnetism submit to the relation

$$e^2/2Gm_e^2 = \chi^{75},$$

where e, $m_e$ - charge and mass of electron, G - gravitational constant.

In fact, that *astonishing* relation points to the *new theoretical* (not experimental) representation of *G gravitational constant* over characteristic constants of *microworld e, $m_e$* - charge and mass of electron.

Simultaneously, it opens the possibility to receive the explicit representation for the $\alpha_g$ gravitational FI constant

$$\alpha_g = 2Gm_e^2/\hbar c \Rightarrow e_g^2/\hbar c$$

and corresponding gravitational charge

$$e_g = (2G)^{1/2}m_e$$

## UNKNOWN POTENTIAL POSSIBLE FI.

Comparison approaches, connected with Standard Model and WU Concept, are not assist to well-being in connection with developed, prevailind representations. Evidently, that value - "wealth" ("power") of FI, Described by $\alpha$ Homology, extremaly more, then "wealth" of FI, proposed by Standard Model. Its are related as $\infty$ :4. What must we do with this "wealth"?

It is so far remain unknown in the modern physical World Picture and, therefore, still non investigated by all available in a modern science intellectual and experimental potential.

It will be reasonable treat with attention to the indications of theory, a'priori not reject a possibility of search of new laws in new areas and, at first, begin purposeful experimental investigations, connected with early unknown fundamental interactions.

**Nearest FI.**